%% file: Amirkhanyan_en.tex
\documentclass[
 aps, pra,
 amsmath,amssymb,
 11pt,
 final,
tightenlines,
 twoside,
 twocolumn,
 nofloats,
nofootinbib,
 superscriptaddress,
showkeys,
showkeywords,
 ]
{revtex4-2}

\usepackage[T2A]{fontenc}
\usepackage[utf8x]{inputenc}
\usepackage[russian,english]{babel}
\usepackage{graphicx}
\usepackage{dcolumn}
\usepackage{bm}
\usepackage{longtable}

\input{maik.rty}

\setcitestyle{authoryear,round,aysep={,},citesep={;}}
\input{sao_cmd_author.tex}

\begin{document}

\selectlanguage{english}

\title{OBJECT S5\,0716+71: FLUX - LINEAR POLARIZATION COUPLING}
\author{\firstname{V.~R.}~\surname{Amirkhanyan}}
 \email{amir@sao.ru}
 \affiliation{Special Astrophysical Observatory of the Russian Academy of Sciences, Nizhnij Arkhyz, 369167 Russia}

\begin{abstract}
The linear polarization observations of S5\,0716+71 carried out by the author in 2019--2021 were continued from December 8, 2021 to March 12, 2022. These observations confirm the author's argument made in 2022 about a periodic dependence of the degree of linear polarization of S5\,0716+71 on its optical flux. The harmonic period varies from 3 to 8~mJy in the 3 to 55~mJy interval.  

 \end{abstract}

  \maketitle
 \section{INTRODUCTION}
The object S5\,0716+714 has been studied by astronomers for over 40 years. It attracted much attention due to its rapid emission variations and its linear polarization. The variations were recorded in a wide interval ranging from radio to gamma rays on time scales from ten minutes to several years. The power spectrum of the variable emission component of S5\,0716+714 is close to flicker noise with no signs of a harmonic component. 

 Such close attention to this object is determined by several factors. Variability, undoubtedly, carries hidden information about the mechanisms of emission and the magnetic field structure of the jet. It is the jet that, when pointed towards the observer, is the main supplier of S5 0716+714 emission according to most researchers.  

This blazar is a rather bright object, circumpolar at our latitudes, which allows us to carry out long series of observations. Note also that observations can be carried out with good accuracy using small instruments whose observation time is less regulated. This paper continues the work of Amirkhanyan (2022), where the author presented the optical linear polarization observation results 
for the well-known object S5\,0716+714 and suggested that the degree of linear polarization of the object depends harmonically on its magnitude. In order to refute or confirm this claim, monitoring and investigation of this object continued.

\section{OBSERVATIONS AND RESULTS}
Observations of S5\,0716+714 were carried out over the period from December 8, 2021 to March 12, 2022 using the same SAO RAS telescope, Zeiss-600, with the same ТАЗ-18 equipment mounted as before. A Savart plate is used as a polariod. In order to determine three Stokes parameters ($I$, $Q$ and $U$) it is sufficient to carry out two exposures in two polariod position angles with an optimal difference of 45~degrees. The polarimeter performs the set number of exposures (up to 65\,536) with a given exposure, changing (after each exposure) the Savart plate position angle by a preset value. The observation techniques and observational material processing methods are the same (see details in Amirkhanyan, 2022).
  In the period from January 17, 2019 to March 12, 2022, 4419 exposures were taken with the Zeiss-600 telescope. In the first days of observations, from January 17 to 20, the exposure time was 60~s. It was then increased to 120~s for all future observations. The observation log is presented in Table~1.

Additionally, Zeiss-1000 (Afanasiev et al., 2021) and BTA SAO RAS (Shablovinskaya and Afanasiev, 2019) S5\,0716+714 observation data obtained by LSPEO SAO RAS staff were used.

The obtained light curves of S5\,0716+714 and standard~5 for the entire period of observations are shown in Fig.~1. To estimate the photometric and polarimetric errors, series of standard~5 observations were used  (see Fig. 1 in Amirkhanyan, 2022),  which are not burdened by variability. The standard shows stable photometry (Villata et al., 1998) and zero polarization. The root-mean-square error of the standard light curve, as is the case in the first paper (Amirkhanyan, 2022), is equal to $0\,.\!\!^{\rm m}006$. The mean polarization level determination error of the zero standard remains at a level of 0.0077, which turned out to be better than the computed 0.0095.
 \renewcommand{\baselinestretch}{0.7}
\begin{table*}
\caption{\space S5\,0716+714 observation log \label{tab1}} \medskip 
\begin{tabular}{c|c|c|c|c||c|c|c|c|c}
\hline
\multicolumn{2}{c|}{Start of observations} & \multicolumn{2}{c|}{End of observations}  & \multirow{2}{*}{$N_{\rm exp}$}&
\multicolumn{2}{c|}{Start of observations} & \multicolumn{2}{c|}{End of observations}  & \multirow{2}{*}{$N_{\rm exp}$}\\
\cline{1-4}
\cline{6-9}
Date  & JD-2450000 & Date  & JD-2450000 & 
&Date  & JD-2450000 & Date  & JD-2450000 & \\
    \hline
2019/01/17    &    8501.4760    &    2019/01/18    &    8501.5428    &    57  &  2020/01/20    &    8869.4771    &    2020/01/20    &    8869.4878    &    6    \\
2019/01/18    &    8502.4294    &    2019/01/19    &    8502.5598    &    90  &  2020/01/24    &    8873.3777    &    2020/01/24    &    8873.4938    &    68   \\
2019/01/20    &    8504.4451    &    2019/01/21    &    8504.5569    &    126 &  2020/01/25    &    8874.3654    &    2020/01/26    &    8874.6468    &    162   \\
2019/01/31    &    8515.4184    &    2019/02/01    &    8515.5100    &    60  &  2020/02/01    &    8881.3617    &    2020/02/01    &    8881.4048    &    30   \\
2019/02/03    &    8518.3889    &    2019/02/04    &    8518.5233    &    90  &  2020/02/19    &    8899.3891    &    2020/02/19    &    8899.4143    &    18   \\
2019/02/04    &    8519.3861    &    2019/02/04    &    8519.4847    &    64  &  2020/02/20    &    8900.3689    &    2020/02/20    &    8900.4568    &    36   \\
2019/02/05    &    8520.4081    &    2019/02/06    &    8520.5253    &    76  &  2020/02/21    &    8901.3712    &    2020/02/20    &    8901.5502    &    120  \\
2019/11/09    &    8796.6103    &    2019/11/09    &    8796.7212    &    67  &  2020/11/20    &    9173.5586    &    2020/11/20    &    9173.7011    &    77   \\
2019/11/10    &    8797.5987    &    2019/11/10    &    8797.7555    &    96  &  2021/01/04    &    9218.5005    &    2021/01/04    &    9218.5391    &    19   \\
2019/11/23    &    8810.6179    &    2019/11/23    &    8810.7187    &    61  &  2021/02/20    &    9266.3250    &    2021/02/20    &    9266.4816    &    93   \\
2019/11/24    &    8811.5691    &    2019/11/24    &    8811.7279    &    100 &  2021/03/02    &    9276.3207    &    2021/03/02    &    9276.4386    &    73    \\
2019/11/25    &    8812.5603    &    2019/11/25    &    8812.7189    &    90  &  2021/12/08    &    9556.5498    &    2021/12/08    &    9556.5906    &    29   \\
2019/11/26    &    8813.5974    &    2019/11/26    &    8813.7135    &    78  &  2021/12/12    &    9560.5209    &    2021/12/12    &    9560.6842    &    101  \\
2019/11/29    &    8816.5438    &    2019/11/29    &    8816.6995    &    100 &  2021/12/29    &    9578.4529    &    2021/12/30    &    9578.5045    &    36    \\
2019/12/05    &    8822.5405    &    2019/12/05    &    8822.6998    &    97  &  2021/12/30    &    9579.4356    &    2021/12/31    &    9579.5810    &    88   \\
2019/12/07    &    8824.5112    &    2019/12/07    &    8824.6966    &    120 &  2022/01/10    &    9589.5011    &    2022/01/10    &    9589.5989    &    61    \\
2019/12/08    &    8825.5060    &    2019/12/08    &    8825.7013    &    118 &  2022/01/17    &    9597.3745    &    2022/01/17    &    9597.4153    &    19    \\
2019/12/08    &    8826.4976    &    2019/12/09    &    8826.6862    &    116 &  2022/01/20    &    9600.3781    &    2022/01/20    &    9600.3994    &    8     \\
2019/12/09    &    8827.4983    &    2019/12/10    &    8827.6895    &    114 &  2022/01/28    &    9608.4493    &    2022/01/29    &    9608.5063    &    39    \\
2019/12/11    &    8828.5072    &    2019/12/11    &    8828.6739    &    102 &  2022/02/03    &    9614.3484    &    2022/02/04    &    9614.5027    &    57    \\
2019/12/18    &    8836.4806    &    2019/12/19    &    8836.5078    &    20  &  2022/02/05    &    9616.4160    &    2022/02/06    &    9616.5166    &    65   \\
2019/12/20    &    8838.4845    &    2019/12/21    &    8838.6556    &    102 &  2022/02/10    &    9621.3041    &    2022/02/10    &    9621.3489    &    32    \\
2019/12/29    &    8846.5173    &    2019/12/29    &    8846.5549    &    24  &  2022/02/11    &    9622.3067    &    2022/02/12    &    9622.5179    &    140  \\
2019/12/31    &    8849.4575    &    2020/01/01    &    8849.5828    &    80  &  2022/02/15    &    9626.2969    &    2022/02/16    &    9626.5353    &    156  \\
2020/01/05    &    8854.4341    &    2020/01/06    &    8854.6168    &    120 &  2022/02/19    &    9630.4648    &    2022/02/20    &    9630.5340    &    48    \\
2020/01/13    &    8862.4518    &    2020/01/14    &    8862.5357    &    52  &  2022/02/28    &    9639.3760    &    2022/02/28    &    9639.4721    &    47   \\
2020/01/14    &    8863.4224    &    2020/01/15    &    8863.6046    &    120 &  2022/03/06    &    9645.3852    &    2022/03/06    &    9645.4578    &    43    \\
2020/01/15    &    8864.3942    &    2020/01/16    &    8864.5753    &    112 &  2022/03/07    &    9646.2905    &    2022/03/07    &    9646.3635    &    50    \\
2020/01/18    &    8867.3833    &    2020/01/19    &    8867.6949    &    183 &  2022/03/12    &    9651.3321    &    2022/03/12    &    9651.4249    &    63    \\
\hline
\end{tabular}
\end{table*}
\renewcommand{\baselinestretch}{1.0} 
\begin{figure*}
\includegraphics[width=0.91\linewidth, angle=180]{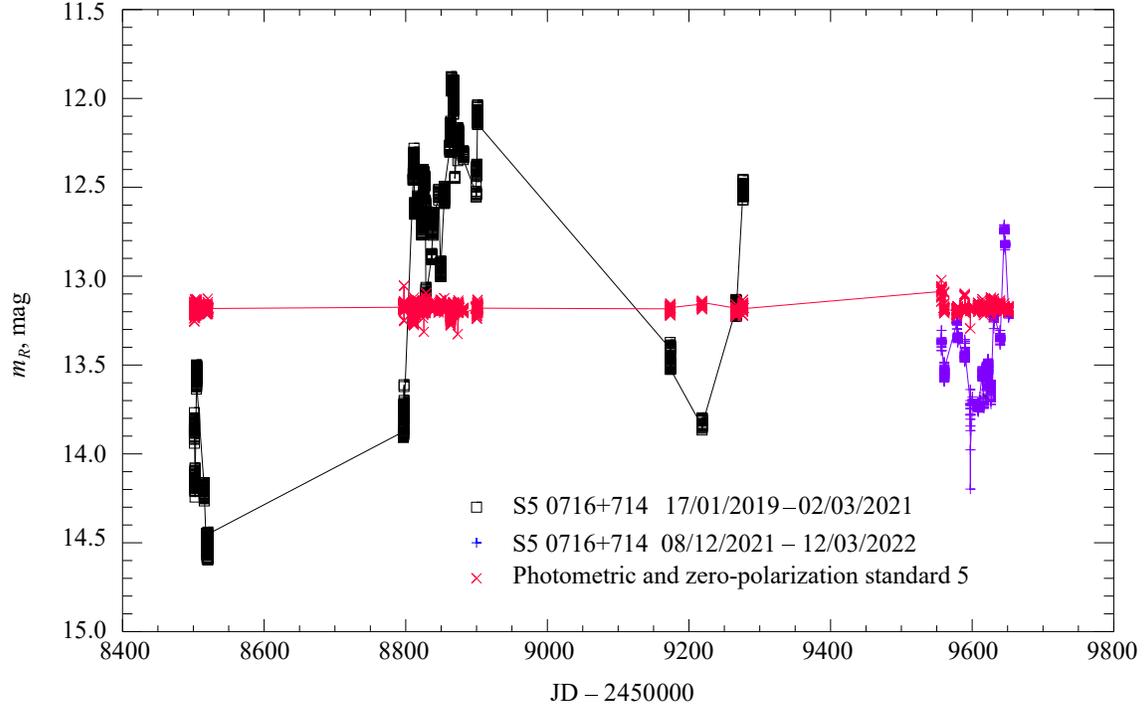} 
 \caption{Light curves of S5\,0716+714 and the closest standard (red line) over the entire period of observations from January 17, 2019 to March 12, 2022. The second series of observations is shown in blue. }
 \end{figure*} 
    
Fig.~2 shows the dependence of the linear polarization level of S5\,0716+714 on its $R$-band flux. For magnitude-to-flux conversion, the calibration from the internet resource\footnotetext{See the information on the website: \url{https://lweb.cfa.harvard.edu/~dfabricant/huchra/ay145/mags.html}}\footnotemark[\value{footnote}] was used. 
The black squares in Fig.~2 show observations from January 17, 2019 to March 2, 2021, and the red squares show those from December 18, 2021 to March 12, 2022. The coinciding fluxes in the two series of observations are in the 10--35~mJy range. As is evident from Fig.~2, the polarization in the two series for close fluxes coincides. In the first paper (Amirkhanyan, 2022) the polarization measured in different epochs coincided in the flux region of more than 40~mJy  (see Fig. 7 in Amirkhanyan, 2022). The connection of the flux with the polarization level of the object can thus be traced in a wide range of observed fluxes.

\begin{figure*} 
 \includegraphics[width=0.91\linewidth, angle=180]{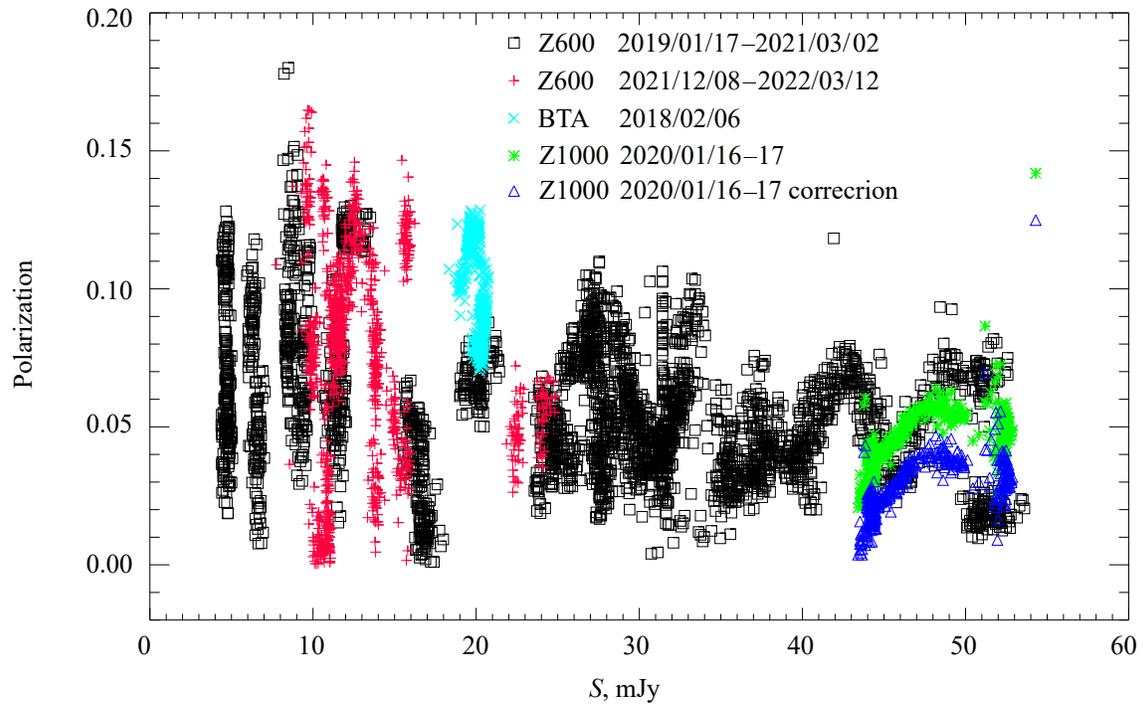}
\caption{``Flux~--- polarization'' dependence for S5\,0716+714. Observations from January 17, 2019 to March 2, 2021 are shown in black. The red color shows observations from December 8, 2021 to March 12, 2022. BTA observations on February 6, 2018 are shown in cyan. Zeiss-1000 observations without correction carried out on January 16-17, 2020 are shown in green, and those with correction are shown in blue. }
\end{figure*}
Besides the Zeiss-600 observations, the plot also shows observations of this object carried out with the following SAO RAS telescopes: BTA, February 6, 2018    (Shablovinskaya and Afanasiev, 2019)  and \mbox{Zeiss-1000}, January 16--17, 2020.  (Afanasiev
et al., 2021). These observations were carried out continuously for over 8~hours with 60~s exposures. As a result,  491 and 461~flux and linear polarization measurements were obtained with the Zeiss-1000 and BTA correspondingly. These data were reduced by the author and are shown in Fig.~2 by cyan (BTA), green, and blue \mbox{(Zeiss-1000)}. It is necessary to clarify the cross referencing procedure of the object fluxes obtained with different telescopes. While the Zeiss-600 observations were carried out in the $R$-band, the \mbox{Zeiss-1000} observations were performed without a filter. The receiving system worked in the 4000--8500~\AA\ wavelength interval. Early morning Zeiss-600 observations on January 16, 2020 and \mbox{Zeiss-1000} partially overlap, which allowed us to match the object fluxes. The object-to-standard~5 flux ratio during the simultaneous observation interval is 1.073 times higher for \mbox{Zeiss-1000} than for Zeiss-600. The  \mbox{Zeiss-1000} photometry was corrected by this value. The polarization degree estimates at the same points are equal to 0.045 for \mbox{Zeiss-1000} and  0.028 for Zeiss-600. The polarization difference is possibly determined by the fact that the receiving band of the Zeiss-1000 equipment is three times wider than the Zeiss-600 band. Fig.~2 shows in green the results without the polarization correction and those corrected by $-$0.017 in blue. The author believes that the Zeiss-600 and Zeiss-1000 data are in a rather good agreement in the  45--55~mJy flux range. The BTA observations were conducted in the $g$-SDSS filter. In order to place the result of these observations on Fig.~2, one must convert the flux to the $R$-band. This can be done if the  $\alpha$ spectral index of the S5\,0716+714 emission is known between the average wavelengths of the $g$- and $R$-bands  ($S\sim \lambda^{-\alpha} $). 

Let us determine the object flux in the $g$-SDSS band (mJy) as described by G.~D.~Wirth\footnotemark[\value{footnote}]: 
 $$
 S_g=3.73\times 10^{6.57-0.4 m_g}. \eqno  (1)
 $$

Amirkhanyan (2006) obtained the dependence of the S5\,0716+714 spectral index on flux $S$ in filters $B$, $V$ and  $I$. Let us use the equation for filter $V$, which is the closest to  $g$-SDSS:
   $$ 
   \alpha=-1.6 + 0.0168 S_V. \eqno(2 )
   $$
 To test this dependence the author used  $V$- and $R$-band photometric series from  Raiteri et al. (2003) to obtain the spectral index as a function of $V$-band flux:
  $$ 
   \alpha=-1.55 + 0.0184 S_V. \eqno(3)
   $$ 
  Both versions give similar spectral index values, which vary in the interval from $-$1.3 to  $-$1.35 
  depending on flux.  Considering the fact that the average wavelengths of the \mbox{$g$-SDSS} and $V$-bands differ only slightly (5200~\AA\  and  5500~\AA\  
     correspondingly), let us change the  $S_V$ fluxes in expressions (2) and (3)  for  $S_g$ obtained with the BTA and compute the $R$-band flux of the object:
         $$
        S_R=S_g(\lambda_R /\lambda_g)^{-\alpha}. \eqno  (4 )\\
        $$        
Here $ \lambda_g=5200$~\AA\  and $\lambda_R=6400$~\AA\ are taken from the data described on the website\footnotemark[\value{footnote}]. 
  The results of these computations are shown in Fig.~2 in cyan. The polarization values were not corrected. The author allows that the position of the cyan fragment may be shifted along the flux axis by several mJy from the real value. The upper limit of the polarization level decreases with increasing flux, which can be described by a simple expression: 
  $$
  P_{\rm max}=0.2(1-0.011S_R). \eqno(5) 
  $$
  At the same time, the boundary of the polarized part of the emission increases with increasing total flux of the object:
  $$
  S_{P_{\rm max}}=0.2(1-0.011S_R)S_R. \eqno(6) 
  $$
  In addition to the monotonic drop in the upper polarization level limit with increasing flux, a non-monotonic periodic component is evident, which repeats itself both in the Zeiss-600 observations and in the Zeiss-1000 and BTA observations.

    \begin{figure*}
 \includegraphics[width=0.79\linewidth, angle=180]{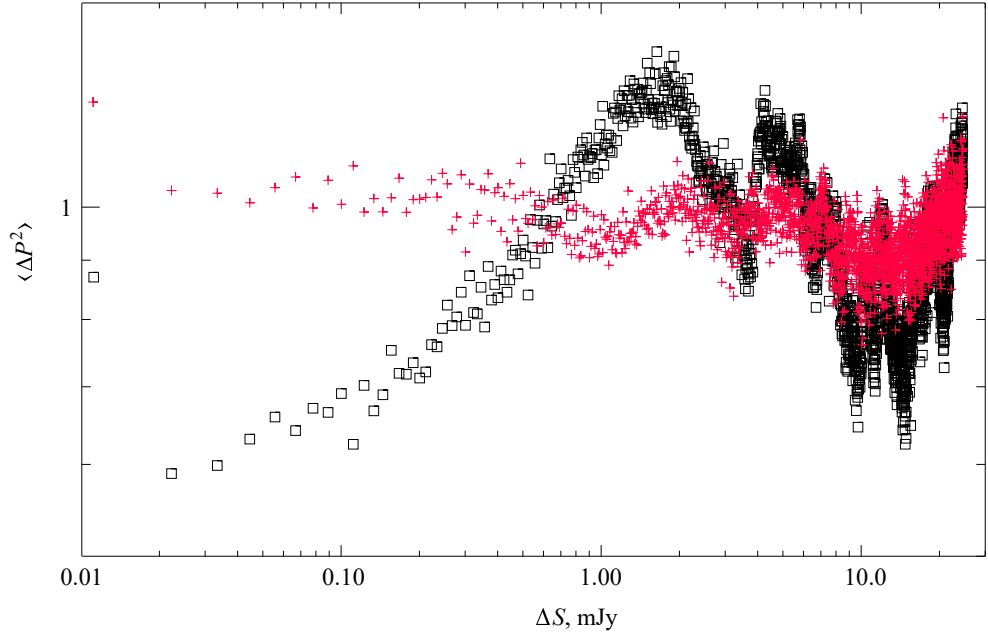} 
 \caption{Black squares: structural function of the ``flux\,---\,polarization'' measurements obtained with \mbox{Zeiss-600}.
 Red crosses: structural function after randomizing the polarization counts.}
  \end{figure*}
      \begin{figure*}
\includegraphics[width=0.8\linewidth, angle=180]{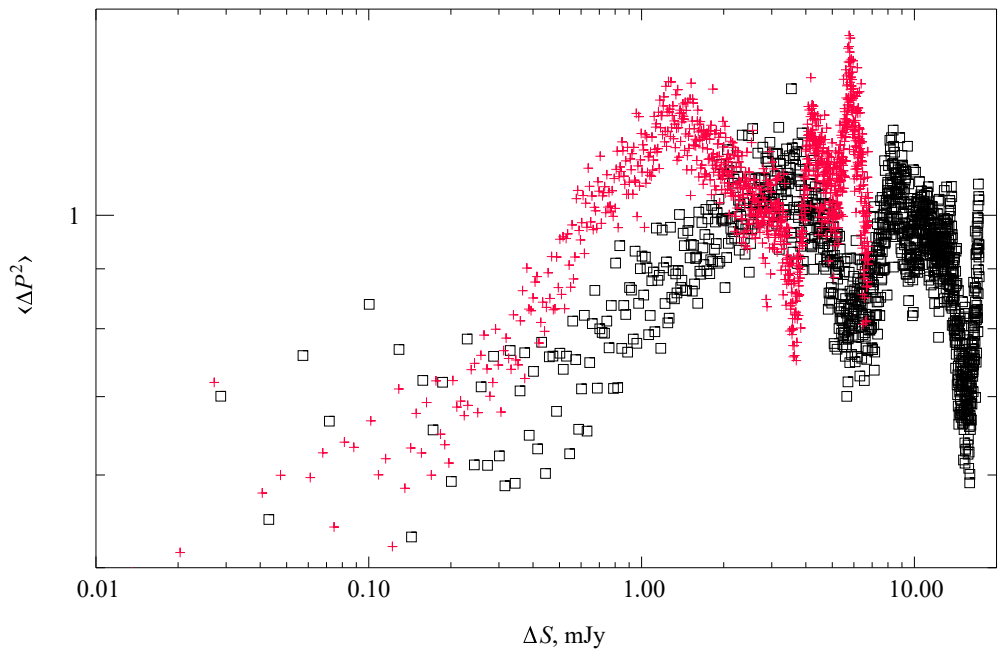} 
  \caption{Structural functions for two sections of the ``flux\,---\,polarization'' array: the black squares show the flux above 18~mJy, the red crosses indicate flux less than 18~mJy.}
   \end{figure*}

    To verify the last statement let us construct the following function~--- the root mean square variation of the polarization difference $\langle\Delta P^2\rangle$ depending on the flux increment $\Delta S_R$, which allows one to estimate the nature of the polarization variations for flux counts of unequal accuracy:
  $$
   \Delta P^2(\Delta S_R)=\langle[P(S_R+\Delta S_R)-P(S_R)]^2\rangle. \eqno(7)
   $$
     For the structural function derivation, only the data from \mbox{Zeiss-600} were used, as the most homogeneous.
    The graph of the structural function (Fig.~3, black squares) shows that the dependence of polarization on flux has a harmonic component. Its period is twice the position of the first maximum, the amplitude of which, ideally, tends to 2., and also twice the difference of the maximum and minimum positions of the graph. 
    \begin{figure*} \includegraphics[width=0.85\linewidth, angle=180]{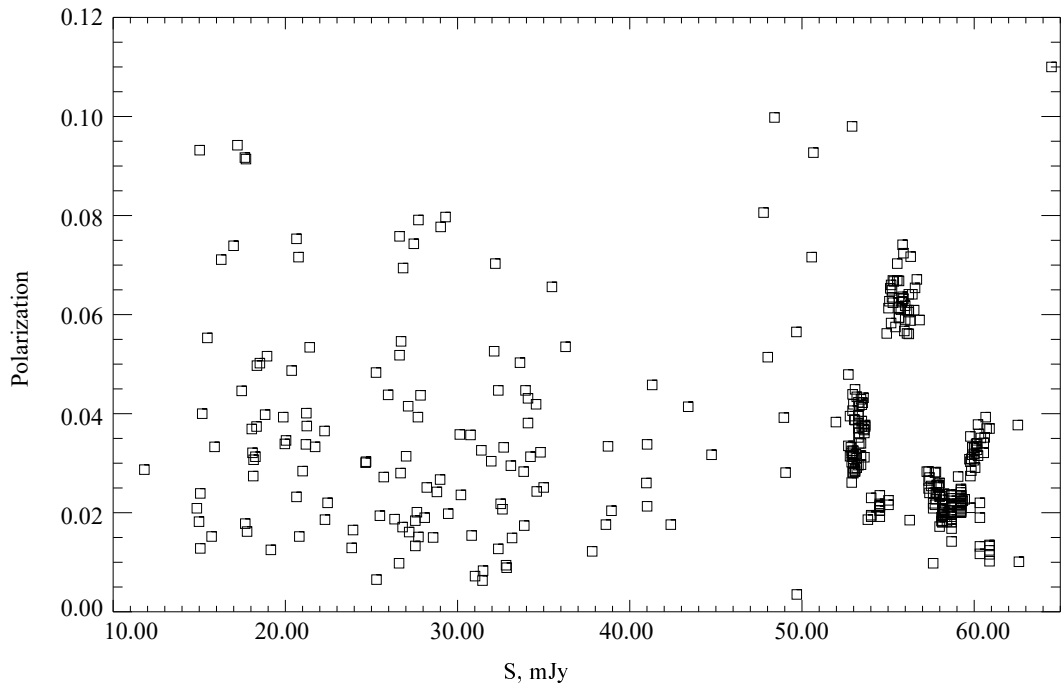} 
   \caption{ ``Flux\,---\,polarization'' dependence based on the data from  Fraija et al. (2017).}
 \end{figure*}

 For comparison, the result of a numeric experiment where random polarization values, distributed according to the Rayleigh law, are substituted for the real ones and shown in the same figure in red. The upper distribution boundary, as is the case in Fig.~2, is given by expression (5).  If the polarization counts $P(S_R)$ are a random process, then the \mbox{$P(S_R+\Delta S_R)-P(S_R)$ } difference is close to stationary  (Rytov et al., 1976). Therefore,   $\langle\Delta P ^2(\Delta S_R)\rangle$ depends only on the flux increment $\Delta S_R$ and changes little, and its error increases with increasing increment due to the decreasing number of averaged counts. Precisely this fact is demonstrated by the red plot. The same structural function for standard~5, whose polarization is determined only by the measurement errors, maintains a constant average value at the level of unity in the entire \mbox{0.01--3}~mJy flux shift range, as is the case with the model function in Fig.~3. Note also an important circumstance~--- the number of averaged counts used in the structural function construction changes (in our case) from 2500 to 1000. Therefore, due to the central limit theorem, the distribution of the sample mean of random numbers, which is the structural function count, degenerates into a normal distribution. The theorem is true for any distribution of the initial series with a finite dispersion. The above allows one to argue that besides the monotonic component, the experimental ``flux~--- polarization'' series for S5\,0716+714 also includes a variable component with a period of about 3.5~mJy.

 The author divided the observations presented in Fig.~2 into two sections~--- from 0 to 18~mJy (2001~counts) and from 18~mJy to 55~mJy (2418 counts). For each section the structural function was constructed. In Fig.~4 they are shown by black squares (18--55~mJy) and red crosses (0--18~mJy). Evidently, a harmonic with a period of 2.5--3.5~mJy dominates in the ``weak'' section, and that with a period of 7--9~mJy, in the ``strong'' section.   

Studying multiple publications dedicated to this object allows one to find papers with the \mbox{``flux\,---\,polarization degree''} dependence derived, or with available observational series on flux and polarization (Ikejiri et al., 2011; Smith et al., 2009;  Larionov et al., 2013; Doroshenko et al., 2017; Ahnen et al., 2018).

Our result is not confirmed in any of these papers. Although the structural function constructed using the materials of the last work shows (if one desires) the presence of a weak harmonic in the ``flux~--- polarization'' dependence. The optical observations were carried out by a wide range of telescopes, including the Crimean AZT-8 and the Saint Petersburg LX-200. At these telescopes, unlike the others, photometric and polarization measurements were made every observing night with a long series of exposures. The author does not exclude the possibility that it is these observations that left their mark on the structural function. 

The author then considered the results of observations of the well-known Lacertae type type objects: BL\,Lac  (Hagen-Thorn et al., 2002; Smith et al.,
2009), 3C\,279  (Kiehlmann et al., 2016),  OJ287  (Villforth et al., 2010),  S4\,0954+65  Morozova
et al., 2014),  Mkr\,421  (Fraija et al., 2017).  Some of these papers discuss a positive or negative correlation between the object flux and polarization degree. Rapid variations in flux and polarization are usually attributed to the superposition of the synchrotron radiation of several groups of relativistic electrons, or their motion through the entangled and regular magnetic fields. No variable component is noted in the ``flux\,--\,polarization'' dependence. Although the author saw an interesting picture in the graph plotted using the data from Fraija et al. (2017) (see Fig.~5). In the part where the flux exceeds 50~mJy, the nature of the graph is close to Fig.~2 and contrasts sharply with the region of weaker fluxes. This may be due to the fact that in the region of strong fluxes, observations were carried out every night from April 13 to 19, 2013 with long series of 60 second exposures, whereas only 1--2 exposures were taken per night in the region of weak fluxes.   

\section{CONCLUSIONS}
Independent observations of the linear polarization of S5\,0716+714 with three SAO RAS telescopes  show that the \mbox{``flux\,--\,polarization''} relation has a harmonic component. Multi-hour series of regular exposures were performed on each night of \mbox{Zeiss-600} observations. Under good weather conditions, up to 100--150 exposures were taken. Over 450~exposures per night were taken with the 6-meter (BTA) and the \mbox{1-meter} (Zeiss-1000) telescopes. Detailed light and polarization curves were obtained, which allowed us to trace the variations of these parameters on scales of hours. We were able to register a harmonic connection of linear polarization with the object flux on a scale of 3--8~mJy. The absence of this effect in both S5\,0716+714 and in several other objects of this type in the works of other authors seems to be due to the limited number of exposures and averaging the results of a night of observations. We shall continue observing blazars in the same regular multi-hour regime. If the effect is confirmed, then its explanation will need to be found that does not contradict the absence of a harmonic flux component, which we have been searching for in such objects for several decades.

\end{document}

%% file: sao_cmd_author.tex
\def\squareforqed{\hbox{\rlap{$\sqcap$}$\sqcup$}}

\def\sq{\ifmmode\squareforqed\else{\unskip\nobreak\hfil
\penalty50\hskip1em\null\nobreak\hfil\squareforqed
\parfillskip=0pt\finalhyphendemerits=0\endgraf}\fi}

\def\utw{\smash{\rlap{\lower5pt\hbox{$\sim$}}}}

\def\udtw{\smash{\rlap{\lower6pt\hbox{$\approx$}}}}

\def\diameter{{\ifmmode\mathchoice
{\ooalign{\hfil\hbox{$\displaystyle/$}\hfil\crcr
{\hbox{$\displaystyle\mathchar"20D$}}}}
{\ooalign{\hfil\hbox{$\textstyle/$}\hfil\crcr
{\hbox{$\textstyle\mathchar"20D$}}}}
{\ooalign{\hfil\hbox{$\scriptstyle/$}\hfil\crcr
{\hbox{$\scriptstyle\mathchar"20D$}}}}
{\ooalign{\hfil\hbox{$\scriptscriptstyle/$}\hfil\crcr
{\hbox{$\scriptscriptstyle\mathchar"20D$}}}}
\else{\ooalign{\hfil/\hfil\crcr\mathhexbox20D}}%
\fi}}

\newcommand{\aap}{Astron. and Astrophys. }
\newcommand{\aas}{Astron. and Astrophys. Suppl. }

\newcommand{\aj}{Astron.~J. }
\renewcommand{\apj}{Astrophys.~J. }
\newcommand{\apjs}{Astrophys.~J. Suppl. }

\newcommand{\mnras}{Monthly Notices Royal Astron. Soc. }

\newcommand{\pasj}{Publ. Astron. Soc. Japan }